\def\nonotes{1}
\documentclass[sigconf]{acmart}

\usepackage{xspace} 
\usepackage{booktabs} 

\usepackage[utf8]{inputenc}
\usepackage[T1]{fontenc}

\usepackage[np,autolanguage]{numprint}
\usepackage{multirow}
\usepackage[normalem]{ulem}

\usepackage{xargs}                      

\ifdefined\nonotes
\usepackage[disable,colorinlistoftodos,textsize=footnotesize]{todonotes}
\else
\usepackage[colorinlistoftodos,textsize=footnotesize]{todonotes}
\fi

\usepackage[colorinlistoftodos,textsize=footnotesize]{todonotes}
\newcommandx{\fixthis}{\todo[linecolor=OliveGreen,backgroundcolor=OliveGreen!25,bordercolor=OliveGreen]{FixThis}\xspace}
\newcommandx{\rediger}{\todo[inline,linecolor=OliveGreen,backgroundcolor=OliveGreen!25,bordercolor=OliveGreen]{Rédiger}\xspace}
\newcommandx{\afaire}[1]{\todo[inline,caption={},linecolor=OliveGreen,backgroundcolor=OliveGreen!25,bordercolor=OliveGreen]{TODO: #1}}
\newcommandx{\unsure}[2][1=]{\todo[inline,caption={},linecolor=red,backgroundcolor=red!25,bordercolor=red,#1]{Vérifier: #2}}
\newcommandx{\change}[2][1=]{\todo[inline,caption={},linecolor=blue,backgroundcolor=blue!25,bordercolor=blue,#1]{Changer: #2}}
\newcommandx{\improvement}[2][1=]{\todo[inline,caption={},linecolor=Plum,backgroundcolor=Plum!25,bordercolor=Plum,#1]{Améliorer: #2}}
\newcommandx{\thiswillnotshow}[2][1=]{\todo[inline,disable,#1]{#2}}

\setcopyright{rightsretained}

\acmConference[RecSys'18]{ACM RecSys conference}{October 2018}{Vancouver, BC.
  Canada}
\acmYear{2018}
\copyrightyear{2018}

\newcommand{\ie}{\textit{i.e.}\xspace}
\newcommand{\xgboost}{\textit{XgBoost}\xspace}

\begin{document}
\title{Movie rating prediction using content-based and link stream features}

\author{Tiphaine Viard}
\affiliation{\institution{CEDRIC, CNAM Paris}
\institution{RIKEN AIP, Tokyo, Japan}}
\email{tiphaine.viard@riken.jp}


\author{Raphaël Fournier-S'niehotta}
\affiliation{\institution{CEDRIC, CNAM Paris}}
\email{fournier@cnam.fr}


\renewcommand{\shortauthors}{}

\begin{abstract}
  While graph-based collaborative filtering recommender systems have been
introduced several years ago, there are still several shortcomings to deal with,
the temporal information being one of the most important. The new link stream paradigm is aiming at extending graphs for correctly modelling the graph dynamics, without losing crucial information.

We investigate the impact of such link stream features for recommender systems. by designing link stream features, that capture the intrinsic structure and dynamics of the data. We show that such features encode a fine-grained and subtle description of the underlying recommender system.
Focusing on a traditional recommender system context, the rating prediction on
the MovieLens20M dataset, we input these features along with some content-based ones into a gradient boosting machine (XGBoost) and show that it outperforms significantly a sole content-based solution. 

These encouraging results call for further exploration of this original modelling and its integration to complete state-of-the-art recommender systems algorithms. Link streams and graphs, as natural visualizations of recommender systems, can offer more interpretability in a time when algorithm transparency is an increasingly important topic of discussion. We also hope to sparkle interesting discussions in the community about the links between link streams and tensor factorization methods: indeed, they are two sides of the same object.

\end{abstract}

%



\maketitle

\section{Introduction}


Collaborative filtering algorithms are at the core of recommender systems research.
They rely on finding \textit{similar users} to the user for whom the recommendation is intended, collecting previous opinions of these users to compute scores for the items the given user has not yet rated, and present the items with the best scores.
The most widespread modelisation of recommender data is in the form of a matrix where the rows represent the users, the columns represent the items, and one element of the matrix indicates the rating the user has given to the item.

This ensemble of users and items may also be seen as a bipartite graph, where nodes represent users and items, and an edge between a user-node and an item-node represents a rating between the user and the item.
Finding similar users in this context is naturally linked to the graph-theoretic notion of \textit{neighborhood}, \ie the user-nodes which share a subset of neighbour item-nodes with a given node.

While recommender systems initially discarded any notion of time, and two ratings given several years apart where considered equal, a body of research has emerged to take this into account.
A common solution is to rely on sequences of user-item matrices, with a time step ($\Delta$): one builds a sequence $\{M^k\}_{k}$ such that for all $k$, $M^k$ is a user-item matrix, and $M^k_{i,j} \ne 0$ indicates that user $i$ has interacted with item $j$ at least once in $[k,k+\Delta[$.
\afaire{cette dernière phrase peut sauter, pour la place}

Studying dynamic graphs traditionally relies on sequences of \textit{snapshot} graphs, similar to the sequences of user-item matrices $M^k$, with the same shortcomings.
To overcome these issues, the complex network community has come up with \textit{link streams}, also called temporal networks or time-varying graphs depending on the context.
The link stream paradigm enables to study jointly the topological structure and the dynamics of interaction streams.
A system where users interact with items over time, like a recommender system scenario, may then be efficiently modelled as a bipartite link stream.

In this paper, we show that modelling a recommender system as a link stream provide descriptors that are relevant to a recommendation task, including in the context of large-scale recommender systems.
We use the interactions between movies and users from the MovieLens 20M dataset, describe it with content-based features as well as link stream-based features, and finally use state-of-the-art machine learning (\xgboost) to learn the recommendation task.
We evaluate the relevance of such link stream features by comparing their performance to a content-based-only baseline, and to state-of-the-art results on this dataset.

The remainder of this paper is organized as follows: in Section~\ref{sec:problem}, we present the recommending context we study~; the bipartite link stream model is formalised in Section~\ref{sec:model}. We detail the features we devised in Section~\ref{sec:features}.
We present our experimental setup and results in Section~\ref{sec:evaluation}.
We discuss related works in Section~\ref{sec:relatedwork}, before concluding the paper with some stimulating perspectives in Section~\ref{sec:conclusion}.


\section{Problem setting}
\label{sec:problem}



Given a user $u$ and a movie $i$, we focus on the task of predicting the rating assigned by $u$ to $i$ on the $0.5$ scale from $0$ to $5$, \ie\ a \emph{regression} task.
Let $U$ be a set of users, and $I$ a set of movies, with $|U|=n$ the number of users and $|I|=m$ the number of movies, and $F$ a set of features such that $|F|=f$.
Our model inputs a $(n\cdot m) \times f$ matrix, and outputs a $n\cdot m$ matrix $P$, where each element $r$ of $P$ is the predicted rating, between $0$ and $5$.

The feature set $F$ is composed of a mixture of numerical and categorical variables describing the dataset, and, for each user $u\in U$ and movie $i\in I$, is comprised of three sets: $F(u)$, the set of user-based features, $F(m)$, the set of movie-based features, and $F(u,i)$, the set of interaction-based features. The contents of each of these sets is discussed below.
    
\section{The bipartite link stream model}
\label{sec:model}

Modelling a recommender system as a \textit{bipartite graph} $G = (U,I,E_G)$ is rather natural: the sets of users $U$ and items $I$ represent the two sets of nodes $U$ and $I$.
The set of edges $E\subseteq U\otimes I$ is composed of interactions between a user and an item, where $\otimes$ is a shorthand notation for a pair of distinct elements~\cite{latapy2017stream}.
All those sets may be completed by weights, for example a rating, or labels, for example a list of timestamps.
However, these solutions are limited by essence; not resorting to weighted or labeled graphs causes important losses of information, while weighted and labeled graphs are complex objects that currently lacks the vast array of algorithms required for social network analysis.

\begin{figure}[!h]
\centering
\includegraphics[width=.7\columnwidth]{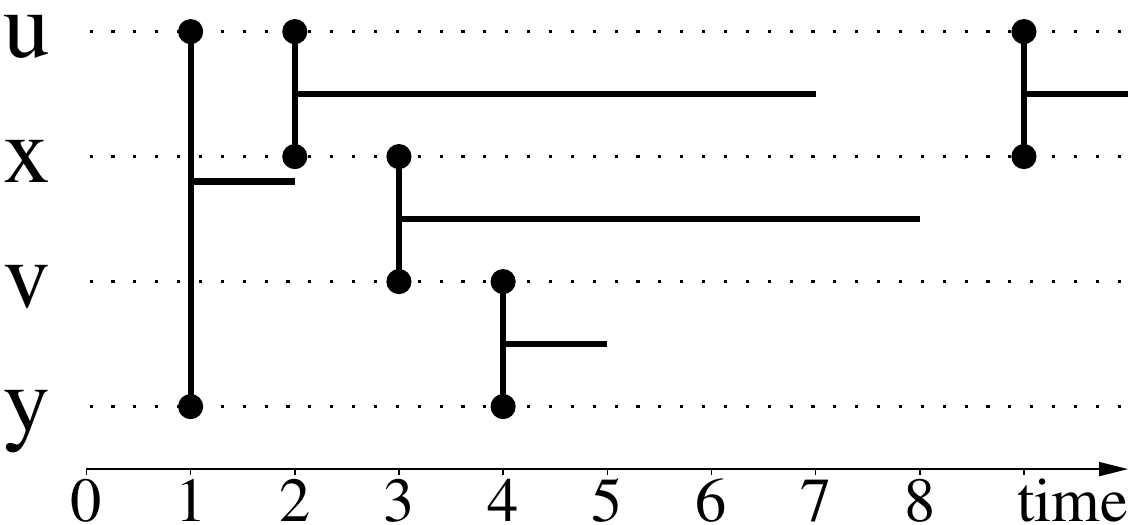}
\caption{
A bipartite link stream $L = (T,U,I,E_L)$ with $T=[0,10]$, $U=\{u,v\}$, $I=\{x,y\}$, and $E_L = ([2,7]\cup[9,10])\times\{ux\} \cup [1,2]\times\{uy\} \cup ([3,8]\times\{vx\} \cup [4,5]\times\{vy\}$. In other words, the links of $L$ are $(2,7,ux)$, $(9,10,ux)$, $(1,2,uy)$, $(3,8,vx)$, and $(4,5,vy)$. We display nodes vertically and time horizontally, each link being represented by a vertical line at its beginning that indicate its extremities, and an horizontal line that represents its duration.
}
\label{fig:linkstream}
\end{figure}

A \textit{bipartite link stream} $L = (T,U,I,E_L)$ is defined by a time span $T$, a set of users $U$, a set of items $I$, and a set of links $E_L \subseteq T \times U \otimes I$, where $\otimes$ is a shorthand notation for a pair of distinct elements~\cite{latapy2017stream}. Nodes $u$ and $i$ are linked at time $t$ if $(t,ui) \in E$. We say that $(b,e,ui)$ is a link of $L$ if $[b,e]$ is a maximal interval of $T$ such that $u$ and $i$ are linked at all $t$ in $[b,e]$. See Figure~\ref{fig:linkstream} for an illustration.

The usual properties of graphs (neighbourhoods, paths, clustering, etc.) have been generalized to link streams, enabling the study of interaction streams with a single modelling structure, and without resorting to snapshots. As with \textit{bipartite graphs}, it is easy to see a recommender system as a \textit{bipartite link stream} $L = (T,U,I,E_L)$.


\section{Dataset and feature engineering}
\label{sec:features}

\subsection{Dataset}

For our evaluation, we focus on the MovieLens 20M dataset, comprising \np{20000263} interactions of \np{138493} users on \np{27278} movies, over the course of 20 years (from January 9\textsuperscript{th}, 1995 to March 31\textsuperscript{st}, 2015). The datasets contains two types of interactions: a user $u$ gives a rating $r\in [0,5]$ to item (movie) $i$ at time $t$, or a user $u$ assigns textual tags to item (movie) $i$ at time $t$. Movies have limited information associated with them, only release year and genres. No demographic information about the users is present, contrary to other MovieLens datasets.

We do not use IMDB identifiers to extract more movie information, which would require NLP\footnote{NLP: Natural Language Processing (tokenisation, lemmatisation)} to come up with good features. Similarly, user-generated tags were only used to augment the number of connections between users and movies.

\subsection{Content-based features}

From the available information in the dataset, we obtained $39$ content-based features. Among them, $19$ were the result of a $n$-out-of-one strategy for the genre of the movies (18 possible genres and a "(no genres listed)" column). Fourteen features code the decade in which the movie was released -- we unsuccessfully experimented with other binning strategies for the release year. The remaining 6 features are the rating mean, median and standard deviation, and the minimum, maximum and number of ratings (normalized). Those last 6 features are computed both for movies and users.

\subsection{Graph and link stream features}

Equipped with the link stream model, we devised $21$ features, some of them being close to what can be found in other graphs models, others being completely original. 
We detail them in the following, relying on two structures: a bipartite link stream $L=(T,U,I,E_L)$ and the bipartite graph it induces, $G=(U,I,E_G)$.

{\bf Neighbourhood-based features.} These features explore the relations between the users and the movies over time.
We say that $u$ is a neighbour of $i$ if there exists at least one interaction between $u$ and $i$, in the dataset.
The degree of a node $u$ in the graph $G$ is simply the number of neighbours of $u$, \ie\ $d_G(u) = |\{i: ui\in E_G \}|$.
This, however, does not take into account the dynamics of a user's neighbourhood; we focus on the neighbourhood of $u\in U \cup I$ at each time $t\in T$: $d_t(u)=|\{i:\exists (t,ui)\in E\}|$.
We describe the evolution of $d_t(u)$ with its mean value, $d(u)=\frac{1}{|T|}\int_t{d_t(u)} dt=\frac{1}{|T|}\sum_{m\in I} |\{(t,ui)\in E_L\}|$, and its maximum value, $\max(d_t(u))$.
Notice that $\max(d_t(u))$ is not necessarily equal to $d_G(u)$.

We also computed the minimum stream degree $\min(d_t(u))$ and the standard deviation of $d_t(u)$, however these two features showed little relevance; we discard them.

  We also compute the {\em assortativity }~\cite{newman2003mixing} of each link $ui \in E_G$, $a(ui) = \frac{\min(d(u), d(i))}{\max(d(u),d(i))}$, which is the ratio between the degrees of the nodes.
    In this context, low values of assortativity typically correspond to famous blockbusters that all users have likely seen.

{\bf Inter contact time features.} To take into account the dynamics of the link stream, we use the sequence of durations between two links involving $u$, defined as follows.
For each user (resp. movie) $u\in U$ (resp. $I$), let $t(u)=(t:(t,ui)\in E_L\cap T\times I \otimes \{u\})$ be the ordered sequence of times at which there is a link involving $u$. 
Then, the inter-event times sequence is the sequence of the differences between two consecutive elements of $t(u)$, \ie\ $\tau(u)= (t_{i+1} - t_i)_{i=0}^{|t(u)|-1}$.
We describe this sequence, for each user (resp. movie), by its maximum $max(\tau(u))$, minimum $min(\tau(u))$, mean $\mu(\tau(u))$ and standard deviation $\sigma(\tau(u))$. 
\afaire{je suis passé de $\max$ à $max$, je trouvais cela plus lisible}

{\bf Clique-based features.}
As an exploratory approach to find clusters of users and items in the bipartite link stream, we rely on cliques in the link stream model, for lack of an established clustering algorithm.
However, enumerating all the maximal cliques is computationally intractable on large data. Plus, some cliques (like stars) do not capture relevant information for recommendation. 
We then use the methodology described in~\cite{viard2018discovering} to sample maximal balanced bipartite cliques, \ie\ cliques involving approximately the same number of users and items.
This kind of object is interesting from a recommender systems point of view: it corresponds to dense subgroups of users all rating a substantial number of items.

Formally, $(U',I',[b,e])$ is a clique in a link stream if all nodes of $U'\subseteq U$ interact with all nodes of $I'\subseteq I$ over $[b,e]\subseteq T$, \ie\ $\forall t\in [b,e], \forall ui\in U' \otimes I', \exists~(t,ui)\in E_L$.
A clique is maximal if it is included in no other.


    From our set of sampled balanced maximal cliques, we computed the following features for each user (resp. movie) $u$:
        \begin{itemize}
            \item The balancedness of the cliques involving $u$:
                $$
                \frac{1}{|C_u|} \sum_{(U',I',[b,e])\in C_u} \frac{\min(|U|, |I|)}{\max(|U|, |I|)}   
                $$
            \item The normalized average duration of the cliques involving $u$: 
                $$\frac{1}{C_u}\sum_{(U', I',[b,e])\in C_u} \frac{|[b,e]|}{|T|}$$
            \item The fraction of cliques containing $u$: 
$$\frac{|\{(U',I',[b,e]) : u\in U\cup I\}|}{|\{(U',I',[b,e]\}|}$$
        \end{itemize}

        where $C_u = |\{(U',I',[b,e]) : u\in U\cup I\}|$ is the number of cliques involving node $u$, a normalizing factor.
        All these features tend to describe the sampled maximal balanced cliques containing $u$.
        Intuitively, nodes belonging to a high fraction of all cliques are typically high-raters.

\section{Evaluation setting and results}
\label{sec:evaluation}

\subsection{Evaluation Metrics}

The main metrics to evaluate the prediction performance of a recommender system are MAE and RMSE~\cite{herlocker2004evaluating}. 
The {\bf Mean Absolute Error (MAE)} is the average error between elements of the ground truth $y$ and the predicted elements $\hat{y}$: 
$$
MAE(\hat{y}, y) = \frac{1}{n}\sum_{i=1}^{n} |\hat{y}_i - y_i|
$$

For a set of $n$ predictions $\hat{y}$ for which the ground truth $y$ is known, the {\bf Root Mean Squared Error} is defined as:
$$
RMSE(\hat{y}, y) = \sqrt{\frac{1}{n}\sum_{i=1}^{n} (\hat{y}_i - y_i)^2}
$$

The MAE is a simple metric that is readily interpretable, but is less sensitive to outliers than the RMSE, for example. 
The RMSE is less easy to interpret, however it offers a good tool of comparison to the state-of-the-art, and is more sensitive to outliers, which is interesting in a prediction context. 

We also report our results for the \textit{NDCG@k} metric, the Normalized Discounted Cumulative Gain for a ranking of $k$ elements. It is formulated as:
$$
NDCG(k) = N\sum_{i=1}^{k} \frac{2^{R(i)}-1}{\log_2(1+i)},
$$
where $R(i)$ is the graded relevance of item $i$, and $N$ is a normalizing factor, such that a perfect ranking has a {\em NDCG} value of $1$.

\subsection{Results}

We perform a 5-fold cross validation on the dataset, and report the results according to our evaluation metrics in Table~\ref{table:results}. The experiment ran for $10$ hours on a machine with 32 8-cores CPUs and 64 GB of RAM.

We tuned \xgboost\ using Bayesian optimization~\cite{snoek2012practical} on the hyperparameter space, and obtain optimal results with deep trees and a small learning rate.
Optimality is reached after a few thousands boosting rounds~\footnote{We use $\alpha=0.3649$, \texttt{eta=0.1}, \texttt{min\_child\_weight=18.6967}, \texttt{colsample\_bytree=0.9112}, $\gamma=0.9930$, \texttt{max\_depth=10} and \texttt{subsample=0.9810} for the solution with link stream features, and $\alpha=0.3649$, \texttt{eta=0.1}, \texttt{min\_child\_weight=18.6967}, \texttt{colsample\_bytree=0.9112}, $\gamma=0.9930$, \texttt{max\_depth=10} and \texttt{subsample=0.9810} for the content-based one.}.


We compare the performance of our algorithm using link streams features along with content-based ones in \xgboost in Table~\ref{table:results}.
We see that adding link stream features lead to significantly better learning performance than using only content-based features.

Our results outperform some of the recent literature~\cite{yao2017efficient}.
To the best of our knowledge, the minimum RMSE obtained on the MovieLens 20M dataset was $0.7652$~\cite{strub2016hybrid} with a deep learning approach.
While we are not there yet, we detail some perspectives in Section~\ref{sec:conclusion} to close this gap.



\begin{figure*}[htbp]
    \begin{tabular}{|c|c|c|c|c|}
\hline
\multirow{2}{*}{Metric} & \multicolumn{2}{c|}{With stream features} & \multicolumn{2}{c|}{Without stream features}\\
& train & test & train & test\\
\hline
MAE & {\bf0.615 ($\pm$ 0.00024)} & {\bf0.63399 ($\pm$ 0.00014)} & 0.63421 ($\pm$ 0.00013) & 0.64277 ($\pm$ 8.6e-05)\\
\hline
RMSE & {\bf0.80256 ($\pm$ 0.00031)} & {\bf0.82913 ($\pm$ 0.0003)} & 0.82682 ($\pm$ 0.00017) & 0.83961 ($\pm$ 0.00025)\\
\hline
NDCG@10 & {\bf0.99283 ($\pm$ 0.014)} & {\bf0.97128 ($\pm$ 0.024)} & 0.98212 ($\pm$ 0.022) & 0.94863 ($\pm$ 0.048)\\
\hline
\end{tabular}

    \caption{$MAE$, $NDCG@10$ and $RMSE$ values the train and test datasets, for our solution and solution without link stream or graph features.}
    \label{table:results}
  \end{figure*}


\subsection{Discussion of feature importance}

\begin{figure}[ht]
    \includegraphics[width=\linewidth]{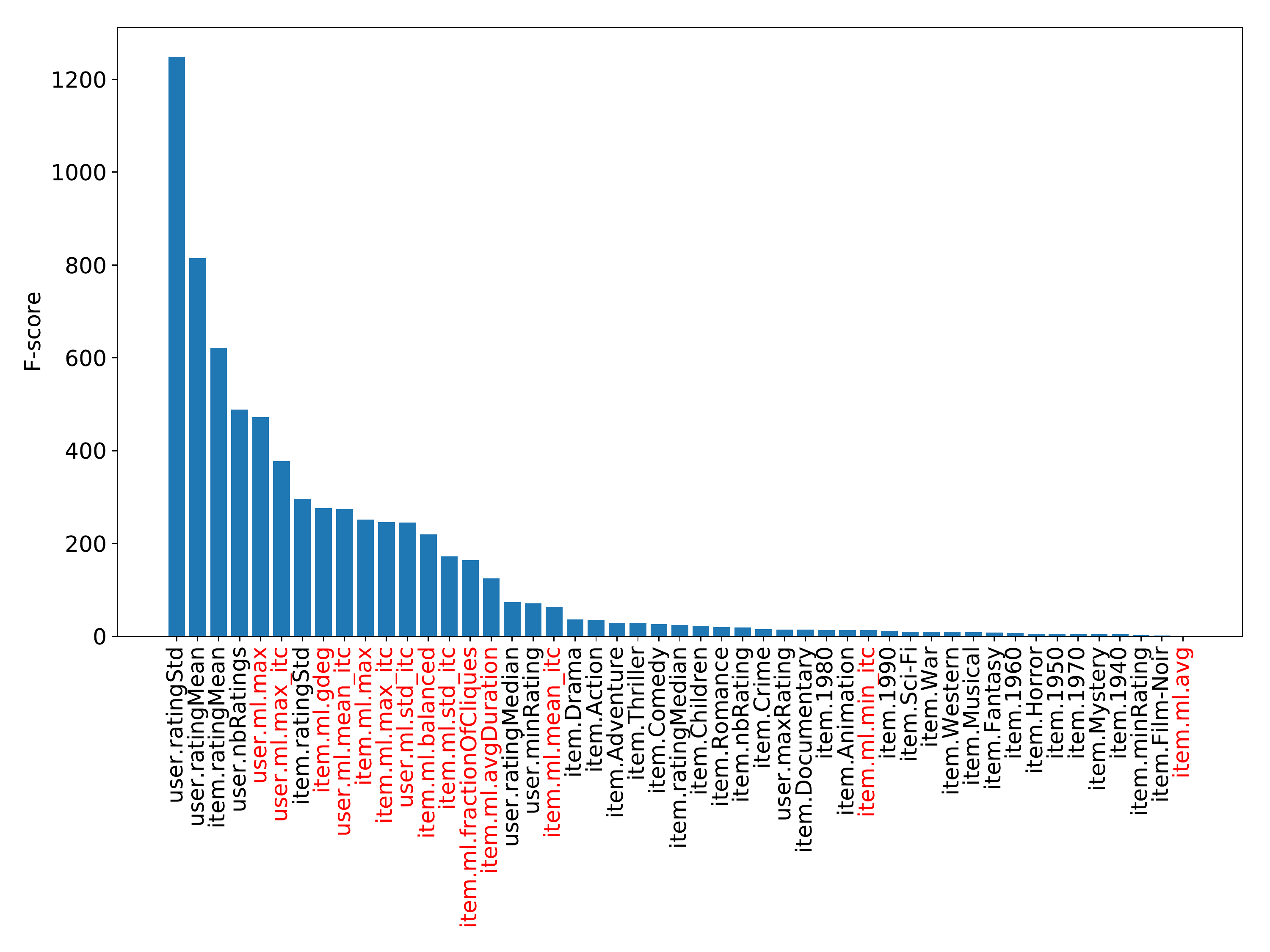}
    \caption{Feature importance as selected by \xgboost. Link stream features are indicated in red. Feature importance is calculated as {\em "the number of times a variable is selected for splitting, weighted by the squared improvement to the model as a result of each split, and averaged over all trees"~\cite{elith2008working}}.}
    \label{fig:feature_importance}
\end{figure}

In addition to the classical performance metrics presented above, we evaluate the descriptiveness of the link stream features we devise. Figure~\ref{fig:feature_importance} shows the relative importance of features as selected by \xgboost, with the link stream features indicated in red.

We can see that the introduced link stream and graph features are commonly used as split points by the boosting algorithm, which supports the claim that such features are very descriptive of the structure and dynamics of the dataset. 
Out of $20$ graph and link stream features, $14$ of them have a non-zero feature importance.
More importantly, the top-$20$ most important splitting features include $13$ graph and link stream features.

This tends to show that link stream features are considered as very relevant descriptors of the underlying recommender system.

\improvement{Peut-être ajouter un inset -- tableau/histogram -- avec seulement les features rouges, ordonnées par feature importance en regard}

\section{Related Work}
\label{sec:relatedwork}

A graph-based vision of collaborative filtering algorithms was introduced, to the best of our knowledge, in 1999 by \citeauthor{aggarwal1999horting}~\cite{aggarwal1999horting}. Since then, most research directions have focused on unipartite graphs gathered in a \textit{social} environment, where links between user- or item-nodes are explicit, such as a trust network~(see \cite{tang2013social} for a review on these \textit{social recommender systems}).

In~\cite{desrosiers2011comprehensive}, \citeauthor{desrosiers2011comprehensive} presented \textit{path-based} methods in a traditional CF setting, which rely on counting shortest paths between nodes to compute their similarity, and \textit{random-walk based} techniques, which evaluate the probability of reaching nodes by a random walk on the graph. 

While finding similar users and grouping them into communities is a vast subject in social network analysis, it has been attempted with limited success in the recommendation context. See for instance~\citeauthor{bernardes2015social} which use the Louvain algorithm in a collaborative filtering framework and obtain state-of-the-art results~\cite{bernardes2015social}. There are currently no community detection algorithm in the link streams framework, which call for theoretical work.

The use of bipartite networks in a recommending framework was first proposed in~\cite{zhou2007bipartite}: the recommendation problem was presented as a \textit{link prediction} problem, a well-studied subject in the complex networks community. The approach consists in an adequate projection of the bipartite network to embed the information available into the weight of a unipartite graph.
While there have been several attempts at defining and finding bipartite communities~\cite{Barber2007,Murata2009,Beckett2016}, several challenges remain before using them in a recommender system.

Temporal dynamics has a high impact on recommender system performance: rating mean may change individually or globally as time elapses~\cite{koren2010collaborative}. However, before 2010 and ~\citeauthor{koren2010collaborative}'s pioneering paper, it was mostly ignored as a research avenue. Since then, ACM RecSys challenge top-3 contestants have successfully incorporated time-based features into a prediction framework~\cite{Pacuk2016}, and a dedicated workshop took place at the ACM RecSys 2017\footnote{\url{https://sites.google.com/edu.haifa.ac.il/tempreasoninginrs/}}. 

Several approaches have been proposed to understand the temporal dynamics of interactions between entities modelled by a graph, see~\cite{masuda2016guidance} for a general survey. In addition to graph snapshots, it has also been proposed to add node and/or edge attributes to enclose temporal information~\cite{batagelj2016algebraic}. While it leads to simple models with the ability to use traditional graph tools, some key concepts like density, centrality or neighborhood are rather overlooked. Link streams were introduced recently to generalize all graph-related concepts for dynamics structures~\cite{latapy2017stream}. The model was used in a network-security study~\cite{viard2018discovering}, while some graph algorithms are progressively adapted to it~\cite{himmel2016}.

\section{Conclusion and perspectives}
\label{sec:conclusion}

\unsure{Doit-on parler de nos essais de classification?} 

We provide a proof-of-concept of incorporating link streams features in a classical recommender system environment. 
We use the large-scale Movielens 20M dataset and obtain a performance slightly below the state-of-the-art, relying on a limited model focusing on link streams features. 
There are several options to tune and improve the features we use to help close the small performance gap, starting with embedding more collaborative features.

Defining and listing communities of users (or of users and items) is a work to be done in the link stream setting, it may prove very useful for recommender systems. 
An equivalent to the graph-based modularity is yet to be devised. 
Computing nodes' betweenness centralities or clustering coefficients should also give precious information on the dynamic of neighbourhoods. 
In a different direction, exploring the impact of the link stream model on time-sensitive recommendations is a question to examine.

We show that link streams-based algorithms may contribute to improving collaborative filtering performance, and that an intuitive underlying model can be called upon to explain why an item was proposed to a user, improving the justifiability. 
As a drawback, there is the newness of the concept, which still lacks some conceptual tools. 
We hope our experiment may prolong the fruitful exchanges of ideas between the recommender system and social network analysis communities.

\afaire{
\begin{itemize}
    \item Discussion: notre approche est bcp plus lente de l'existant (Xgboost tourne en plusieurs heures); je pense que ça va nous être reproché, on devrait en parler dans la discussion, en insistant sur le côté exploratoire (contrairement à 2 décennies d'optimisations en CF), et le côté intuitivité.
\end{itemize}
}

\ifdefined\notex
\else
\bibliographystyle{ACM-Reference-Format}
\bibliography{./bibliography.bib}
\fi

\end{document}